\documentclass[dvips]{article}
\usepackage{icrctc07}

\newcommand{\micro}{\mbox{\usefont{U}{eur}{m}{n}\char22}}

\title{Discovery of very high energy gamma-ray emission in the W~28 (G6.4$-$0.1) region, and multiwavelength comparisons}
\shorttitle{VHE emission from the W~28 region}
\authors{G. Rowell$^{1\dagger}$, E. Brion$^{2\dagger}$, O. Reimer$^{3\dagger}$, 
  Y. Moriguchi$^{4}$, Y. Fukui$^{4}$, A. Djannati-Ata\"i$^{5\dagger}$, S. Funk$^{3\dagger}$}
\shortauthors{G. Rowell et al.}
\afiliations{$^1$ School of Chemistry \& Physics, University of Adelaide, Adelaide 5005, Australia\\
  $^2$ DAPNIA/DSM/CEA, CE Saclay, F-91191 Gif-sur-Yvette, Cedex, France\\
  $^3$ Stanford University, HEPL \& KIPAC, Stanford, CA 94305-4085, USA\\
  $^4$ Department of Astrophysics, Nagoya University, Chikusa-ku, Nagoya 464-8602, Japan\\
  $^5$ APC, 11 Place Marcelin Berthelot, F-75231 Paris Cedex 05, France \\
  $\dagger$ for the H.E.S.S. Collaboration {\tt www.mpi-hd.mpg.de/hfm/HESS}}

\abstract{
  H.E.S.S. observations of the old-age ($>$10$^4$~yr; $\sim 0.5^\circ$ diameter) composite supernova remnant (SNR) W~28
  reveal very high energy (VHE) $\gamma$-ray emission situated at its northeastern and southern boundaries. The
  northeastern VHE source (HESS~J1801$-$233) is in an area where W~28 is interacting with a dense molecular cloud, 
  containing OH masers, local radio and X-ray peaks. The southern VHE sources (HESS~J1800$-$240 with components labelled
  A, B and C) are found in a region occupied by several HII regions, including the ultracompact HII region W~28A2. 
  Our analysis of NANTEN CO data reveals a dense molecular cloud enveloping this southern region, and our reanalysis
  of EGRET data reveals MeV/GeV emission centred on HESS~J1801$-$233 and the northeastern interaction region.
}

\email{growell@physics.adelaide.edu.au}

\begin{document}
\maketitle

\section{Introduction \& H.E.S.S. Results}
The study of shell-type SNRs at $\gamma$-ray energies is motivated
by the idea that they are the dominant sites of hadronic Galactic cosmic-ray (CR)
acceleration to energies approaching the \emph{knee} ($\sim 10^{15}$~eV) and beyond, e.g. \cite{Ginzburg:1}. 
CRs are then accelerated via the diffusive shock acceleration (DSA) process 
(eg. \cite{Bell:1,Blandford:2}).
Gamma-ray production from the interaction of these CRs with ambient 
matter and/or electromagnetic fields is a tracer of such particle acceleration, 
and establishing the hadronic or electronic nature of the parent CRs in any $\gamma$-ray source is a key
issue. 
Already, two shell-type SNRs, RX~J1713.7$-$3946 and RX~J0852.0$-$4622, exhibit shell-like morphology in VHE $\gamma$-rays 
\cite{HESS_RXJ1713_II,HESS_VelaJnr_II,HESS_RXJ1713_III} to 20~TeV and above.
Although a hadronic origin of the VHE $\gamma$-ray emission is highly likely in the above cases, an electronic origin  
is not ruled out.

W~28 (G6.4$-$0.1) is a composite or mixed-morphology
SNR, with dimensions 50$^\prime$x45$^\prime$ and an estimated distance between 1.8 and 3.3~kpc 
(eg. \cite{Goudis:1,Lozinskaya:1}). 
It is an old-age SNR (age 3.5$\times 10^4$ to 15$\times 10^4$~yr \cite{Kaspi:1}), thought to have entered its 
radiative phase of evolution \cite{Lozinskaya:1}.
The shell-like radio emission \cite{Long:1,Dubner:1} peaks at the northern and northeastern
boundaries where interaction with a molecular cloud \cite{Wootten:1} is established \cite{Reach:1,Arikawa:1}.
The X-ray emission, which overall is well-explained by a thermal model, peaks in the SNR centre but has local enhancements 
in the northeastern SNR/molecular cloud interaction region \cite{Rho:2}. 
Additional SNRs in the vicinity of W~28 have also been identified: G6.67$-$0.42 and G7.06$-$0.12 \cite{Yusef:1}.
The pulsar PSR~J1801$-$23
with 
spin-down luminosity $\dot{E} \sim 6.2\times 10^{34}$ erg~s$^{-1}$ and distance $d\geq9.4$~kpc \cite{Claussen:3},
is at the northern radio edge.
                                   
Given its interaction with a molecular cloud, W~28 is an ideal target for VHE observations. 
This interaction is traced by the high concentration of 1720~MHz OH masers \cite{Frail:2,Claussen:1,Claussen:2}, 
and also the location of very high-density ($n>10^3$~cm$^{-3}$) shocked gas \cite{Arikawa:1,Reach:1}.  
Previous observations of the W~28 region at VHE energies by the CANGAROO-I telescope revealed no evidence for such emission 
\cite{Rowell:1} from this and nearby regions.

\begin{figure*}[th]
  \centering
  \hbox{
    \begin{minipage}{0.55\textwidth}
      \includegraphics[width=\textwidth]{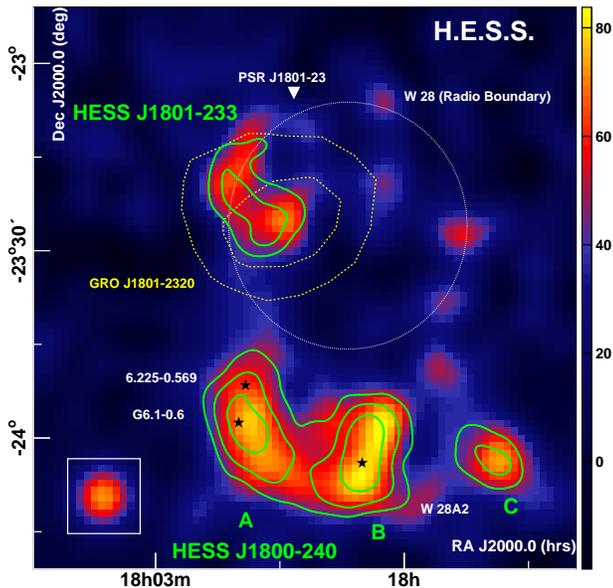}
    \end{minipage}
    \begin{minipage}{0.45\textwidth}    
      \caption{H.E.S.S. VHE $\gamma$-ray excess counts, corrected for 
	exposure and Gaussian smoothed (with 4.2$^\prime$ std. dev.). 
	Solid green contours represent excess  
	significance levels of 4, 5, and 6$\sigma$, for an integrating radius $\theta$=0.1$^\circ$. 
	The VHE sources HESS~J1801$-$233 
	and a complex of sources HESS~J1800-240 (A, B \& C) are indicated.
	The thin-dashed circle depicts the approximate radio boundary of the SNR W~28 \cite{Dubner:1,Brogan:1}.
	Additional objects indicated are: HII regions (black stars); W~28A2, {G6.1$-$0.6} 
	{6.225$-$0.569}; 
	The 68\% and 95\% location contours (thick-dashed yellow lines) of the $E>100$~MeV EGRET source {GRO~J1801$-$2320};
	the pulsar {PSR~J1801$-$23} (white triangle). The inset depicts a pointlike source under identical analysis and smoothing 
	as for the main image.}
      \label{fig:tevskymap}
    \end{minipage}
  }
\end{figure*}

The High Energy Stereoscopic System (H.E.S.S.:  see \cite{Hinton:1} for details and performance) has observed 
the W~28 region over the 2004, 2005 and 2006 seasons.
After quality selection, a total of $\sim$42~hr observations were available for analysis. 
Data were analysed using the moment-based Hillas analysis procedure employing {\em hard cuts} (image size $>$200~p.e.), 
the same used in the analysis of the  inner Galactic Plane Scan datasets \cite{HESS_GalScan,HESS_GalScan_II}. An energy threshold of
$\sim 320$~GeV results from this analysis.
The VHE $\gamma$-ray image in Fig.~\ref{fig:tevskymap} shows that two source of 
VHE $\gamma$-ray emission are located at the northeastern and southern boundaries of W~28.
The VHE sources are labelled HESS~J1801$-$233 and HESS~J1801$-$240 where the latter can be further subdivided into
three components A, B, and C. The excess significances of both sources exceed $\sim$8$\sigma$ after integrating events within
their fitted, arcminute-scale sizes. 
Similar results were also obtained using an alternative analysis \cite{Mathieu:1}.

\section{W~28: The Multiwavelength View}

We have revisited EGRET MeV/GeV data, including data from the CGRO observation cycles (OC) 1 to 6,
which slightly expands on the dataset of the 3rd EGRET catalogue (using OCs 1 to 4; \cite{Hartman:1}, 
revealing the source 3EG~J1800$-$2338. We find a pointlike $E>100$~MeV source in the W~28 region, labelled GRO~J1801$-$2320
in Fig~\ref{fig:tevskymap}. 
The 68\% and 95\% location contours of GRO~J1801$-$2320 match well the location of HESS~J1801$-$233. However we cannot 
rule out a connection to HESS~J1800$-$240 due to the degree-scale EGRET PSF.

\begin{figure*}[h]
  \centering
  \hbox{
      \includegraphics[width=0.5\textwidth]{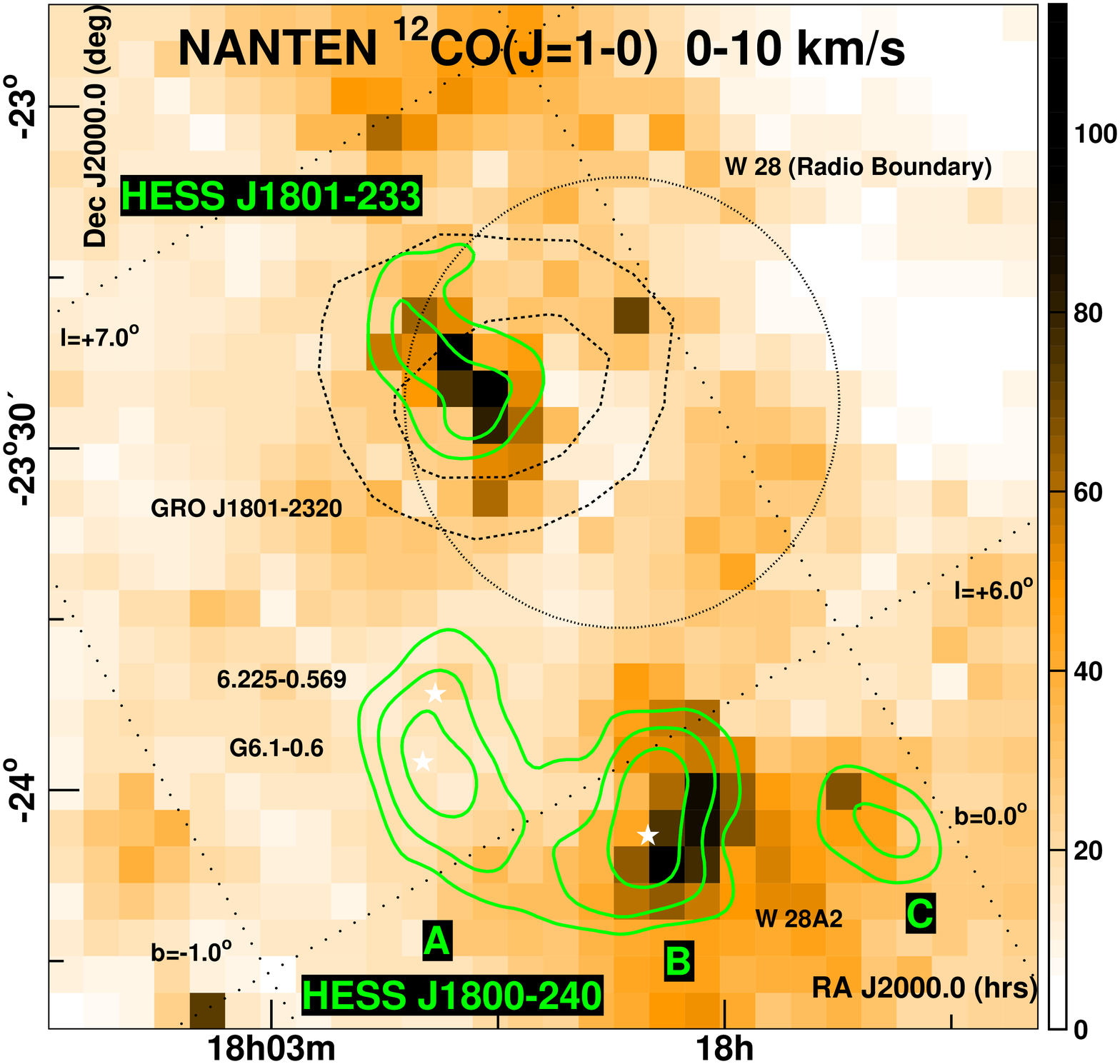}
	\includegraphics[width=0.5\textwidth]{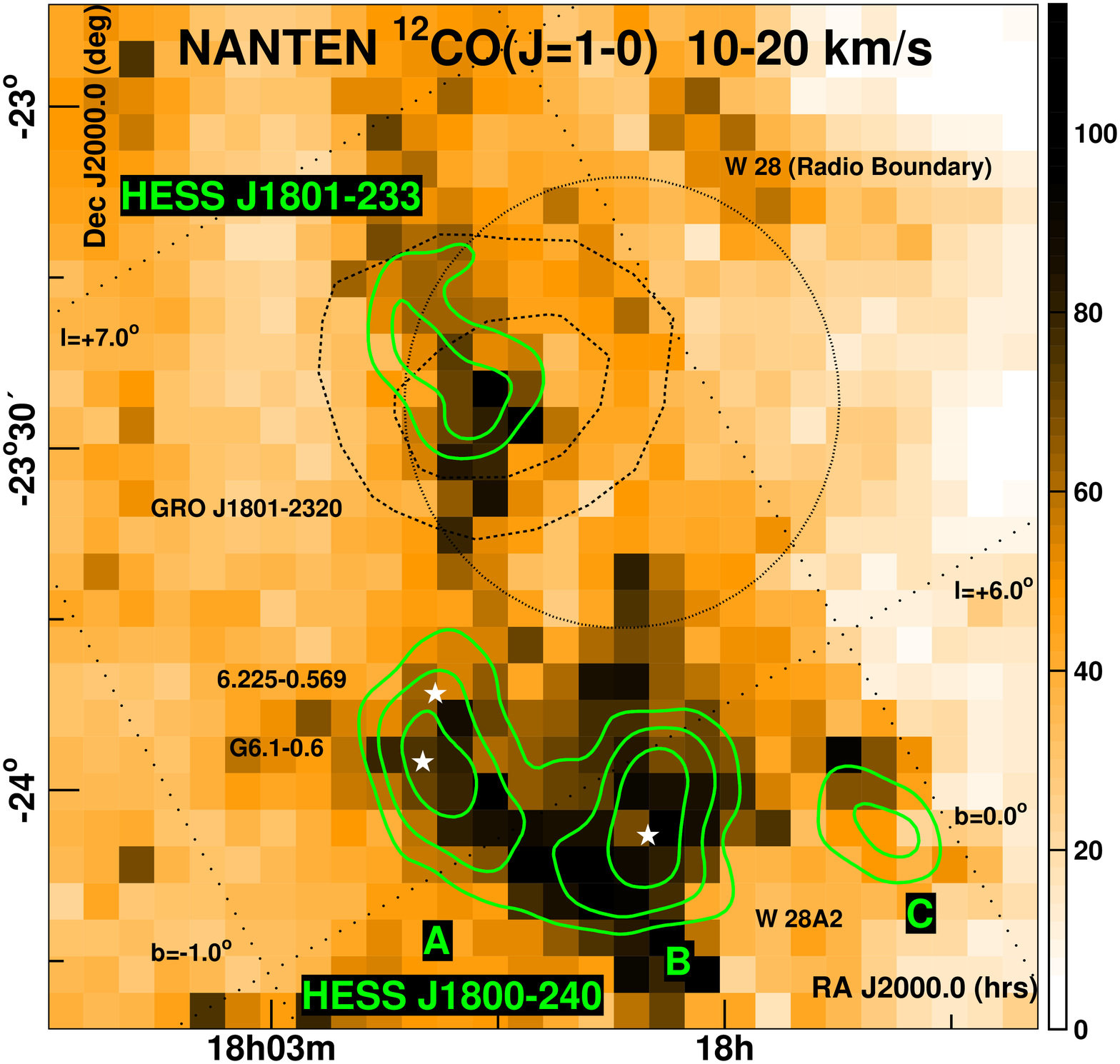}
  }
  \caption{{\bf Left:} NANTEN $^{12}$CO(J=1-0) image 
    (linear scale in K~km~s$^{-1}$) for $V_{\rm LSR}$=0 to 10~km~s$^{-1}$ with VHE $\gamma$-ray significance 
    contours overlaid (green) --- levels 4,5,6$\sigma$ and other features as in Fig.~\ref{fig:tevskymap}. 
    {\bf Right:}  
    NANTEN $^{12}$CO(J=1-0) image for $V_{\rm LSR}$=10 to 20~km~s$^{-1}$ (linear scale and same maximum as for left panel).}
  \label{fig:co_tev}
\end{figure*}

Fig.~\ref{fig:co_tev} presents $^{12}$CO ($J$=1--0) observations from the NANTEN \cite{Mizuno:1} Galactic Plane survey 
\cite{Matsunaga:1} covering the line-of-sight velocity ranges $V_{\rm LSR}$= 0 to 10~km~s$^{-1}$ and 10 to 20~km~s$^{-1}$.
These ranges represent distances 0 to $\sim$2.5~kpc and 2.5 to $\sim$4~kpc respectively and encompass the
distance estimates for W~28. We cannot rule out however, distances $\sim$4~kpc for the  $V_{\rm LSR}>$10~km~s$^{-1}$ cloud components.
It is clear that molecular clouds coincide well with the VHE sources. The northeastern cloud  $V_{\rm LSR}<$10~km~s$^{-1}$ component near HESS~J1801$-$233, is already
well-studied \cite{Reach:1,Arikawa:1}. Contributions from the $V_{\rm LSR}>$10~km~s$^{-1}$ cloud components are also likely.
The molecular cloud to the south of W~28 coincides well with HESS~J1800$-$240 and its three VHE components. 
The $V_{\rm LSR}<$10~km~s$^{-1}$ component of
this cloud coincides well with  HESS~J1800$-$240B, and may represent the dense molecular matter surrounding the ultra-compact HII region W~28A2.
This cloud also extends to $V_{\rm LSR}\sim$20~km~s$^{-1}$ and thus, similar to HESS~J1801$-$233, the total VHE emission in  HESS~J1800$-$240 may result
from several molecular cloud components in projection.  

\begin{figure*}[bh]
  \centering
  \hbox{
    \includegraphics[width=0.5\textwidth]{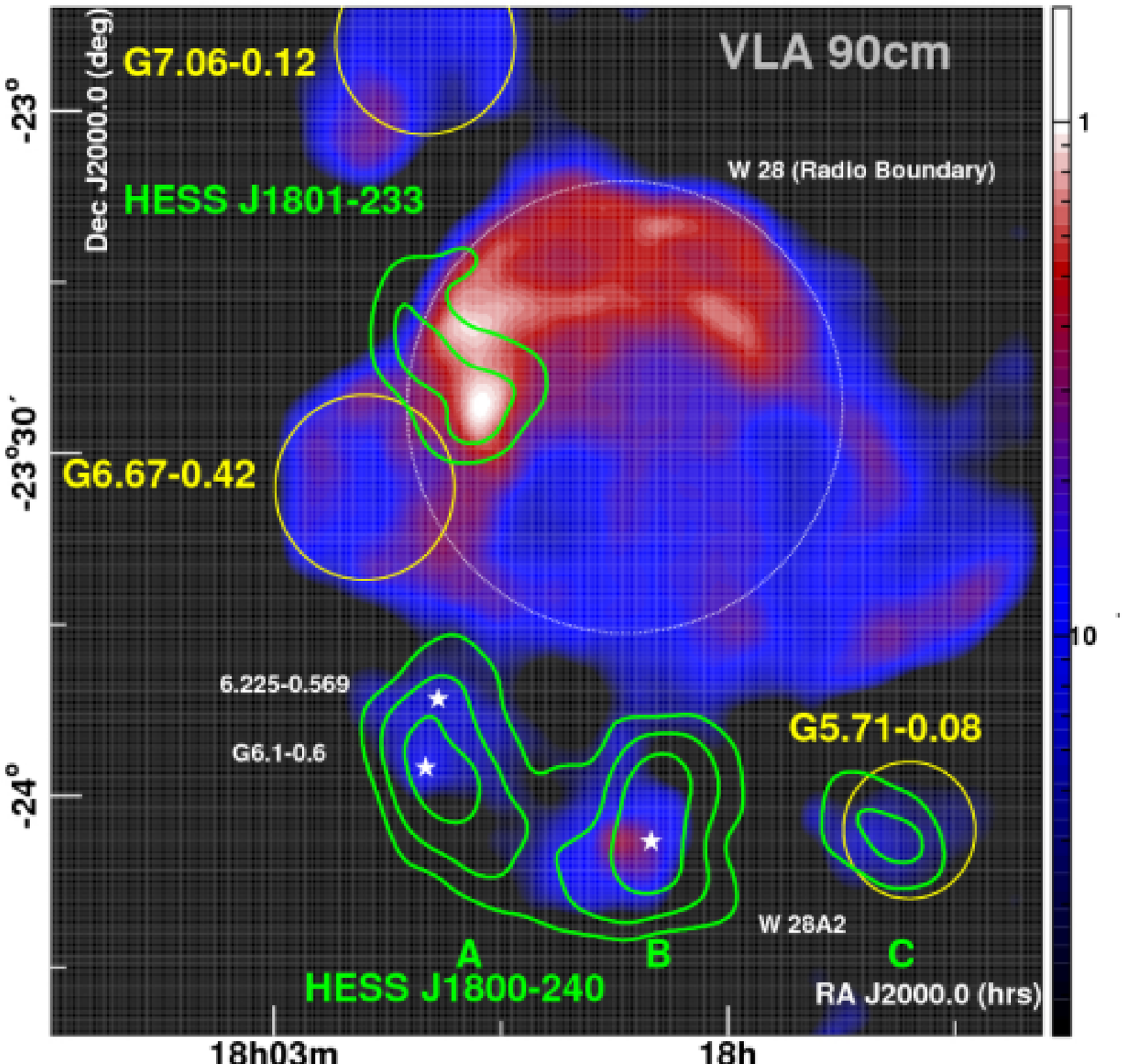}
    \includegraphics[width=0.5\textwidth]{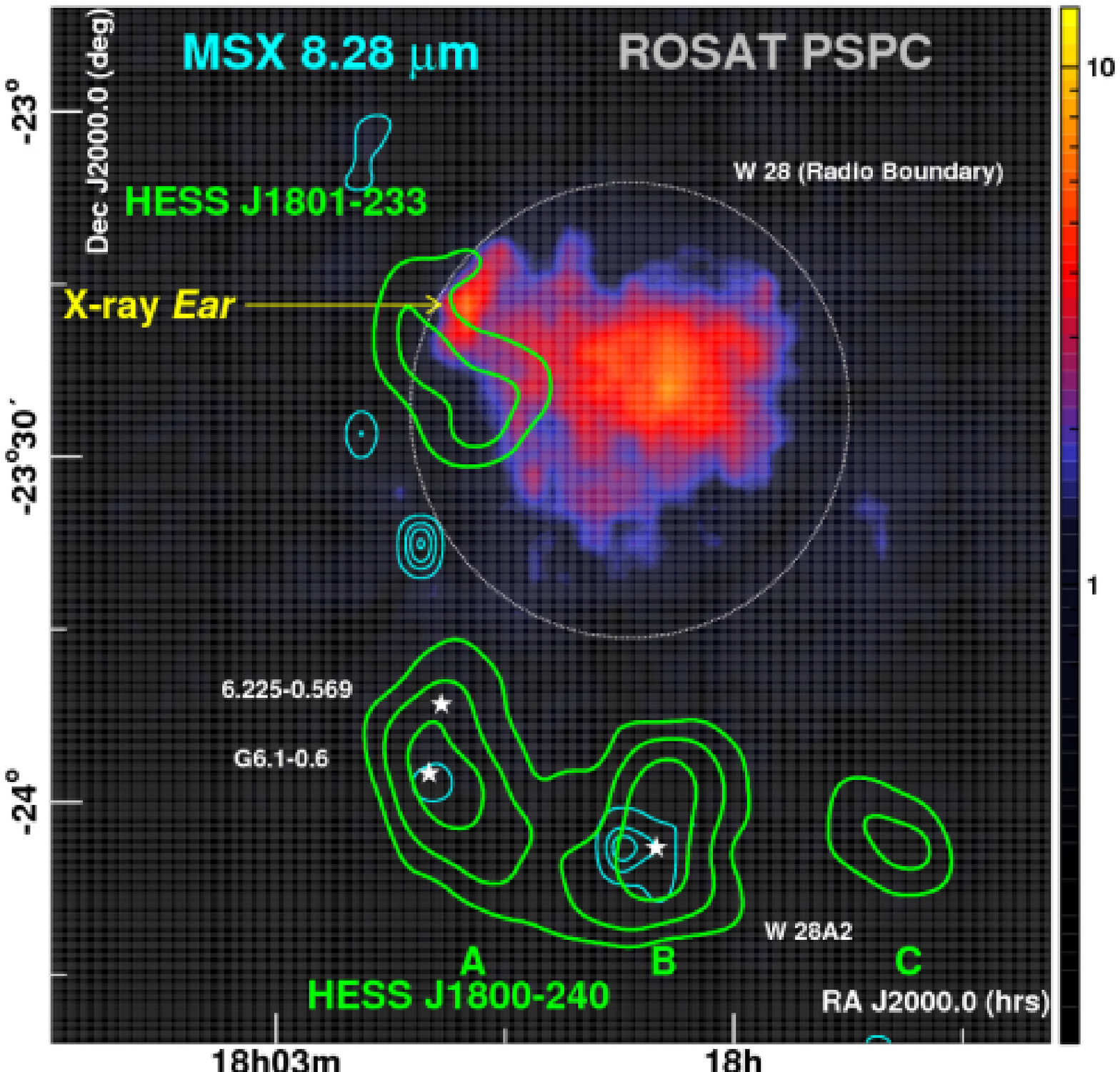}
  }
  \caption{{\bf Left:} VLA 90cm radio image \cite{Brogan:1} in Jy~beam$^{-1}$. 
    The VHE significance contours (green) from 
    Fig.~\ref{fig:tevskymap} are overlaid along with the HII regions (white stars) and the additional SNRs and SNR candidates 
    (with yellow circles indicating their location and approximate dimensions) discussed in text.
    {\bf Right:} 
    ROSAT PSPC image --- 0.5 to 2.4~keV (smoothed counts per bin \cite{Rho:2}). Overlaid are contours 
    (cyan --- 10 linear levels up to 5$\times 10^{-4}$~W~m$^{-2}$~sr$^{-1}$) from the MSX~8.28~\micro m image. Other contours and
    objects are as for the left panel. The X-ray {\em Ear} representing a peak at the northeastern edge is indicated.}
   \label{fig:mwl}
\end{figure*} 

Fig.~\ref{fig:mwl} compares the radio (left panel --- VLA 90~cm \cite{Brogan:1}), infrared and X-ray views (right panel 
MSX 8.28~\micro m and ROSAT PSPC 0.5 to 2.4~keV \cite{Rho:2}) 
with the VHE results. HESS~J1801$-$233 overlaps the northeastern shell of the SNR, coinciding with a strong peak in the 
90~cm continuum emission.
Additional SNRs G6.67$-$0.42 and G7.06$-$0.12 \cite{Yusef:1,Helfand:1} 
are indicated.
The non-thermal radio arc G5.71$-$0.08 is a SNR candidate \cite{Brogan:1}, and is possibly a counterpart to HESS~J1801$-$240C.
The distances to G6.67$-$0.42 and G5.71$-$0.08 are presently unknown. The unusual, ultracompact HII region W~28A2, 
is positioned within $0.1^\circ$ of the centroid of HESS~J1800$-$240B. W~28A2, at a distance $d\sim$2~kpc, 
exhibits energetic bipolar molecular outflows \cite{Harvey:1,Acord:1,Sollins:1} and may therefore be an energy source for 
particle acceleration in the region. The other HII regions 
G6.1$-$0.6 \cite{Kuchar:1} and 6.225$-$0.569 \cite{Lockman:1} are also associated with radio emission.

The X-ray morphology (Fig.~\ref{fig:mwl} right panel) shows the central concentration of
X-ray emission. A local X-ray peak or {\em Ear} is seen at the northeastern W~28 boundary.
The HII regions, W~28A2 and G6.1$-$0.6 
are prominent in the MSX 8.28~\micro m image (Fig.~\ref{fig:mwl} right panel), indicating that 
a high concentration of heated dust still surrounds these very young stellar objects.

\section{Discussion and Conclusions}
H.E.S.S. and NANTEN observations reveal VHE emission in the W~28 region spatially coincident with
molecular clouds. The VHE emission and molecular clouds are found at the northeastern boundary, 
and $\sim 0.5^\circ$ south of W~28 respectively. 
The SNR W~28 may be a source of power for the VHE sources,
although there are additional potential particle accelerators in the region such as other SNR/SNR-candidates,
HII regions and open clusters.   
Further details concerning these results and discussion are presented in \cite{HESS_W28}.

\section{Acknowledgements}

{\renewcommand{\baselinestretch}{-0.5}
\scriptsize
The support of the Namibian authorities and of the University of Namibia\\[-1.5mm]
in facilitating the construction and operation of H.E.S.S. is gratefully\\[-1.5mm]
acknowledged, as is the support by the German Ministry for Education \\[-1.5mm]
and Research (BMBF), the Max Planck Society, the French Ministry \\[-1.5mm]
for Research, the CNRS-IN2P3 and the Astroparticle Interdisciplinary  \\[-1.5mm]
Programme of the CNRS, the U.K. Particle Physics and Astronomy  \\[-1.5mm]
Research Council (PPARC), the IPNP of the Charles University,  \\[-1.5mm]
the Polish Ministry of Science and Higher Education, the South  \\[-1.5mm]
African Department of Science and Technology and National Research  \\[-1.5mm]
Foundation, and by the University of Namibia. We appreciate the  \\[-1.5mm]
excellent work of the technical support staff in Berlin, Durham,  \\[-1.5mm]
Hamburg, Heidelberg, Palaiseau, Paris, Saclay, and in Namibia in the \\[-1.5mm]
construction and operation of the equipment. The NANTEN project is \\[-1.5mm] 
financially supported from JSPS (Japan Society for the Promotion of \\[-1.5mm]
Science) Core-to-Core Program, MEXT Grant-in-Aid for Scientific \\[-1.5mm] 
Research on Priority Areas, and SORST-JST (Solution Oriented \\[-1.5mm]
Research for Science and Technology: Japan Science and Technology \\[-1.5mm]
Agency). We also thank Crystal Brogan for the VLA 90~cm image.}


\scriptsize

\bibliographystyle{plain}
\bibliography{icrc0129}

\end{document}